\documentclass[aip,rsi,reprint,floatfix]{revtex4-1}
\usepackage{graphicx}
\usepackage{amsmath,amssymb,tabularx,color}

\begin{document}

\title{Full-Wave Feasibility Study of Anti-Radar Diagnostic of Magnetic Field 
Based on O-X Mode Conversion and Oblique Reflectometry Imaging}

\pacs{52.35.Hr, 52.70.Gw, 52.55.-s, 07.60.Hv}

\author{Orso Meneghini}
\affiliation{General Atomics, San Diego, California, USA}

\author{Francesco A.~Volpe}
\email[Corresponding author: ]{fvolpe@columbia.edu}
\affiliation{Dept of Applied Physics and Applied Mathematics, 
Columbia University, New York, New York USA}

\begin{abstract}
An innovative millimeter wave diagnostic is proposed to measure the
local magnetic field and edge current as a function of the minor
radius in the tokamak pedestal region. The idea is to identify the direction of 
minimum reflectivity at the O-mode cutoff layer. Correspondingly, the 
transmissivity due to O-X mode conversion is maximum. That 
direction, and the angular map of reflectivity around it, contain information 
on the magnetic field vector $\bf B$ at the cutoff layer. Probing the plasma  
with different wave frequencies provides the radial profile of $\bf B$. 
Full-wave finite-element simulations are presented here in 2D slab geometry. 
Modeling 
confirms the existence of a minimum in reflectivity that depends on the
magnetic field at the cutoff, as expected from mode conversion
physics, giving confidence in the feasibility of the diagnostic. 
The proposed reflectometric approach is expected to yield superior 
signal-to-noise ratio and to access wider ranges of density and magnetic field, 
compared with related radiometric techniques that require the plasma 
to emit Electron Bernstein Waves. 
Due to computational limitations, frequencies of 10-20 GHz were 
considered in this initial study. 
Frequencies above the edge electron-cyclotron frequency 
($f>$28 GHz here) would be preferable for the experiment, because the upper 
hybrid resonance and right cutoff would lie in the plasma, and would 
help separate the O-mode of interest from spurious X-waves.  
\end{abstract}

\maketitle

\section{Introduction and Physical Principle}
Internal measurements of magnetic field and current density 
in a tokamak are important but notoriously difficult 
\cite{TFR, Soltwisch, Wolf, Gentle, Donne}. 
The present numerical work addresses the
feasibility of an innovative millimeter wave diagnostic, which is
expected to measure the local magnetic field vector $\bf{B}$ as a
function of the minor radius in the pedestal region.
As such, it could complement Motional Stark Effect polarimetry 
\cite{MSE,Wroblevski}, 
characterized by good signal-to-noise in the plasma core.

\begin{figure*}[t]
  \includegraphics[width=0.9\textwidth]{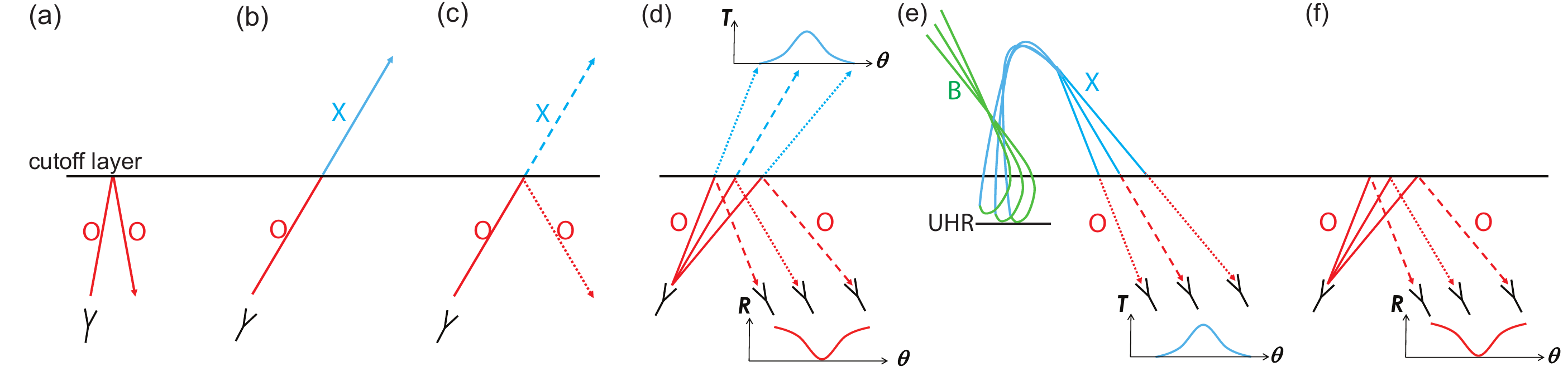}
  \caption{
Physical principle, single ray: (a) cutoff layer ($\omega=\omega_{pe}$) is fully 
reflective for a perpendicular O-mode beam (red), but (b) 
fully transmittive 
for an oblique O-mode of special inclination, dependent  
on local $|{\bf B}|$. Full transmission is due to full conversion into 
X-mode\cite{Mjolus}. (c) Oblique beam of
near-optimal inclination is partly reflected, partly transmitted. 
Bundle of rays: (d) A wide-angle beam
experiences a peak in O-mode transmissivity $T$  
and consequently a hole in O-mode reflectivity $R$. The
anisotropy of R depends on the local vector ${\bf B}$, not solely on 
$|{\bf B}|$. (e) Angular peaks of BXO conversion efficiency have been 
observed in plasmas emitting EBWs \cite{Volpe2010,Shevchenko2011,Vann}. 
(f) A peak in OX conversion
efficiency of an externally launched O-mode and, consequently, a hole in 
O-mode reflectivity, are expected in any device where reflectometry is possible.
Here dashed and dotted lines denote respectively rays of low and high 
intensity, due to partial transmission or reflection.}
\label{fig:princ}
\end{figure*}

\begin{figure}[t]
  \includegraphics[width=0.25\textwidth]{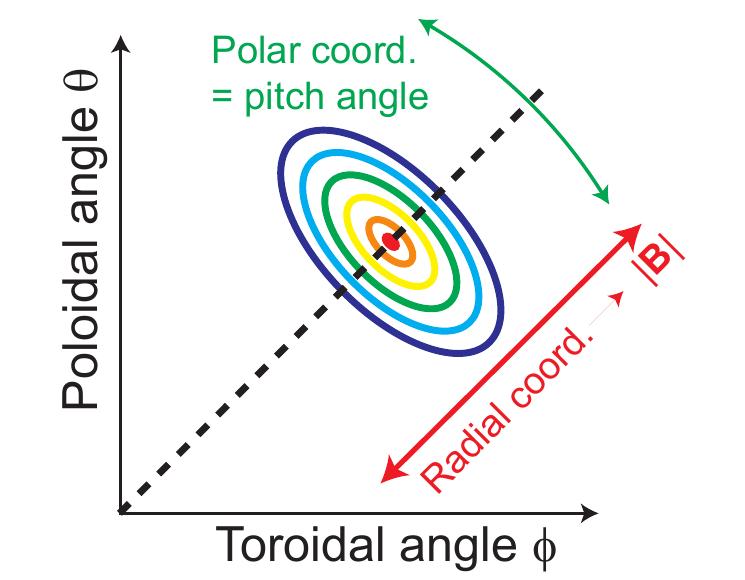}
  \caption{Position and orientation of reflectivity contours 
(plotted ass functions of view angles) are indicative of ${\bf B}$ 
strength and direction at O-mode cutoff, respectively.
}
\label{fig:cont}
\end{figure}

The physical principle, illustrated in Figs.\ref{fig:princ},\ref{fig:cont}, 
is as follows. An angularly broad mm-wave beam
of ordinary (O) polarization is obliquely injected in the magnetized
plasma; at the O-mode cutoff layer, part of the incident wave converts in the 
extraordinary (X) mode with conversion efficiency \cite{Mjolus}

\vskip -0.25cm
\begin{equation}
C=\exp\left\{ -\pi k_0 L \sqrt{\frac{Y}{2}} \left[ 2 (1+Y)
(N_{z,opt}-N_z)^2 +N_y^2 \right] \right\} 
\label{eq:refl}
\end{equation}

and transmits through the cutoff layer. 
Here $L$ is the length scale of density non-uniformity and 
$Y=\Omega_e/\omega$ the dimensionless magnetic field at the cutoff 
location $X=1$, where $X=\omega_{pe}^2/\omega^2$ is the dimensionless density 
and $\Omega_e$, $\omega_{pe}$ and $\omega$ are respectively the electron 
cyclotron, plasma and wave frequencies. The conversion efficiency is maximum 
for incidence in the plane spanned by the density gradient (assumed in the 
$x$ direction) and magnetic field (pointing in the $z$ direction) with 
an optimal $z$ component of the refractive index, $N_{z,opt}=\sqrt{Y/Y+1}$, 
with $Y$ evaluated at the $X=1$ location. 

The rest of the wave keeps the original 
O-mode polarization and is reflected by the O-mode cutoff with 
angle-dependent reflectivity $1-C$. 
From Eq.\ref{eq:refl} it follows that 
$1-C$ contains information on the magnetic field strength (through $Y$). 
It also contains information on the field orientation, because the plane $y-z$ 
is spanned by the density gradient (basically, the minor radius direction) 
and the magnetic field, therefore its inclination in the laboratory frame is 
the same as the magnetic pitch angle. Both pieces of information are 
evaluated at the cutoff layer for the frequency considered. 
That is, the diagnostic is internal and local. 
The angle-dependent reflectivity can be measured using a single launcher 
and an array of receivers. In some sense the diagnostic can be considered 
an ``anti-radar'' that identifies and characterizes minima, rather than 
maxima, of reflectivity. 
Finally, {\em frequency}-resolved measurements of reflectivity 
provide {\em radially} resolved measurements of magnetic field strength and 
pitch angle.

A related, earlier technique relies on partial transmission, though
the O-mode cutoff, of internally emitted electron Bernstein (B) waves
(EBW) undergoing BXO mode
conversion \cite{Volpe2010,Shevchenko2011,Vann}. However, not all fusion
plasmas are over-dense EBW emitters. The new technique proposed here
overcomes this limitation and is applicable whenever reflectometry is 
applicable. Furthermore, the signal to noise ratio is higher, by simply 
adopting a sufficiently intense reflectometric source.  
The spatial and temporal resolution are  
comparable with other reflectometric imaging systems 
\cite{MRI-LHD,MRI-KSTAR,MRI-DIII-D}. 


\section{Full-wave slab model}

To validate this physical principle, we performed preliminary
simulations using the COMSOL Multiphysics code and its RF module 
\cite{COMSOL}. The 
code uses Finite Element Method to solve
arbitrary sets of partial differential equations.
The RF module enables
solution of Maxwell's equations in media characterized by
anisotropic complex dielectric tensors with spatial variations in 1D,
2D and 3D. This software has been used and benchmarked for solving the
problem of antenna-plasma wave coupling and wave propagation in cold
plasmas, as reported in Ref. \cite{Shiraiwa2010}. 
Warm plasma effects 
would be needed to model the BXO conversion used in radiometric techniques 
\cite{Volpe2010,Shevchenko2011,Vann},  
but not for the OX conversion invoked here.

\begin{figure}
  \includegraphics[width=7cm]{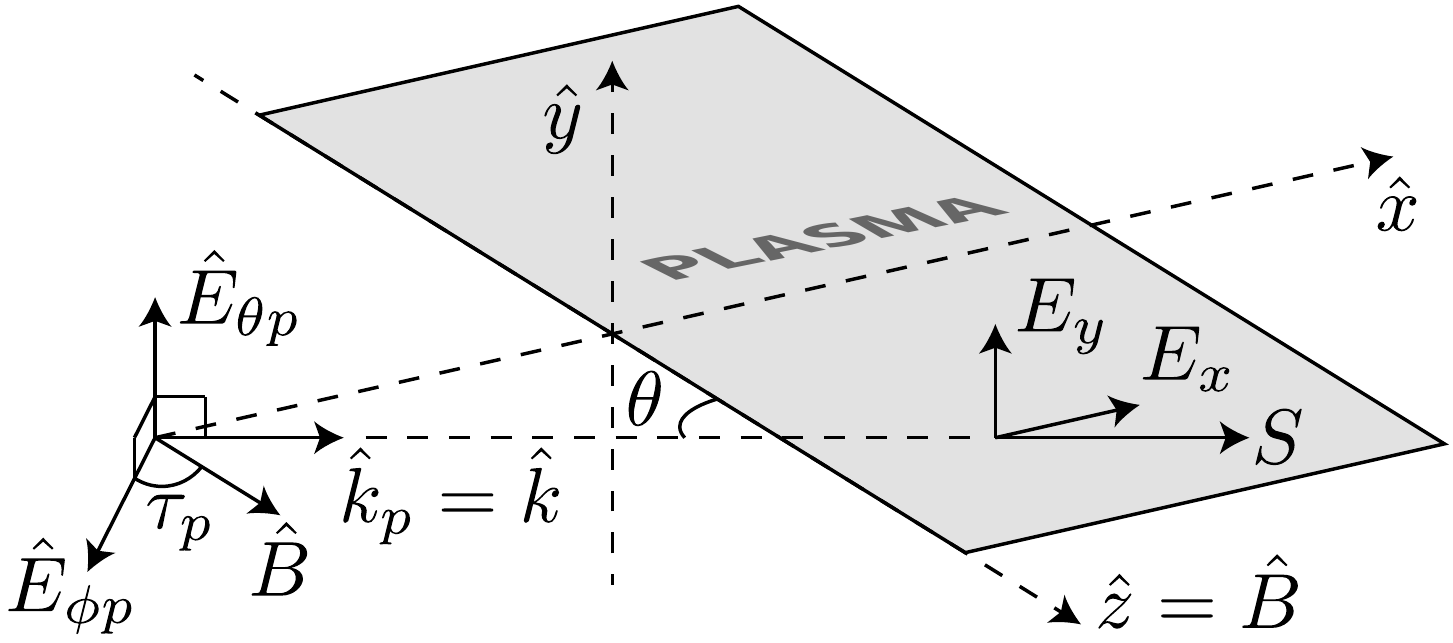}
  \caption{Definition of a set of Cartesian coordinates $\hat x,\hat y, \hat z$ 
in the plasma slab and of an additional set of coordinates, 
rotated to align one of its axes to the wave-launch direction.}
\label{fig:geo}
\end{figure}

\begin{figure}
  \includegraphics[clip=true, trim=400 0 0 0, width=6cm]
                  {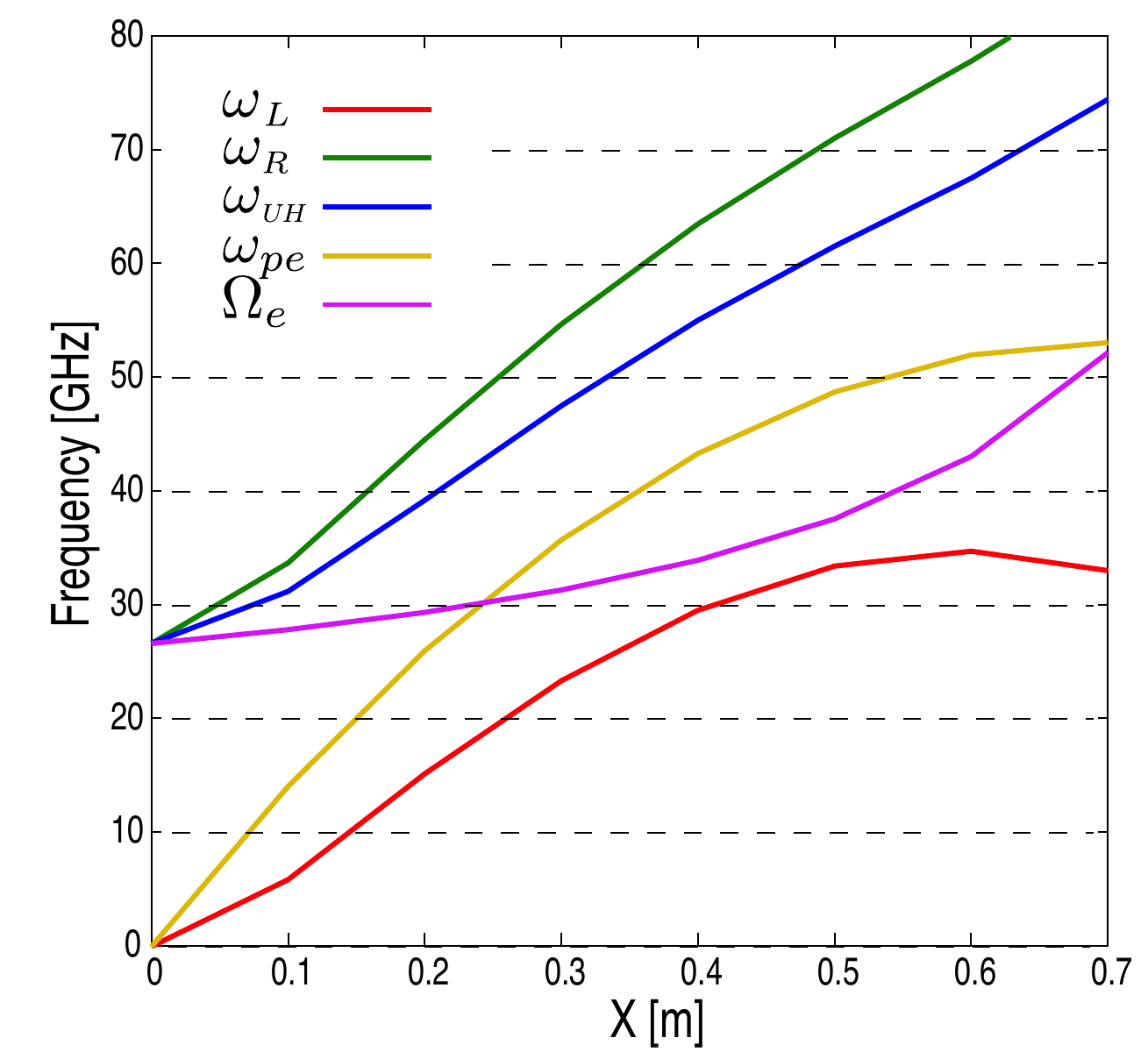}
  \caption{Cutoff and resonant frequencies as a function of the 
  $\hat x$ coordinate.}
\label{fig:res_cut}
\end{figure}

Figure \ref{fig:geo} illustrates the slab geometry chosen
for this study. 
Notice the two sets of Cartesian coordinates. 
All plasma parameters vary only in the $x$ direction, 
playing the role of the radial coordinate. The magnetic field points in the 
$\hat z$ direction, and the wave vector $\bf k$ forms an angle $\theta$ 
relative to it. 
$E_{\theta p}$ and $E_{\phi p}$ are wave field components parallel and
perpendicular to the plane spanned by $\hat x$ and $\hat z$.
From the cold plasma dispersion relation, it follows that the ellipticity of 
the O-mode polarization is given by \cite{Jeong2006}:

\vskip -0.25cm
\begin{equation}\frac{E_{\phi p}}{E_{\theta p}} = \left[ \frac{2 i \sin \tau_p} {Y \cos^2 \tau_p + \sqrt{(Y \cos^2 \tau_p)^2 + 4\sin^2 \tau_p} }\right]
\label{eq:pol}
\end{equation}

where $\tau_p$  is the angle that $E_{\phi p}$ forms relative to the magnetic 
field or, equivalently, relative to $\hat z$. 

In the slab in Fig.\ref{fig:geo}, 
$x$=0 corresponds to the plasma edge on the low-field side, 
and $x$=0.7 m corresponds 
to the plasma center. The magnetic field evaluates 1.9 T at the plasma 
center, and decays like $1/(R+a-x)$, where $R$=0.78 m and $a$=0.7 m.  
The electron density is modeled as $n_e=n_{e0} \left[1 -(1-x/a)^2 \right]^2$, 
with $n_{e0} = 3.5\times10^{19}$ m$^{-3}$. 


Fig.\ref{fig:res_cut} shows the corresponding profiles of 
left cutoff $\omega_L$, right cutoff $\omega_R$, 
upper hybrid resonance $\omega_{UH}$, electron plasma frequency 
$\omega_{pe}$ and electron cyclotron frequency $\Omega_e$. 
The figure refers to perpendicular launch ($\theta=90^o$). 
It should be noted that the radial locations
satisfying $\omega=\omega_{L}$ and $\omega=\omega_{R}$ vary with $\theta$, 
as it will be clearly visible in Fig.~\ref{fig:fwrf}. 

The minimum requirement for this technique is that the O-mode cutoff 
layer defined by $\omega=\omega_{pe}$ lies in the plasma for the 
$\omega$ of choice, so that the O-X conversion can occur. 
For the frequencies adopted here ($\omega /2\pi$=10-20 GHz), 
a wave launched perpendicularly to the plasma at $x$=0 encounters the 
plasma cutoff and the left cutoff (yellow and red in Fig.\ref{fig:res_cut}). 
At higher frequencies, $\omega/2\pi > \Omega_e (x=0)/2\pi = $28 GHz, the wave 
will cross the $\omega=\omega_{R}$ X-mode cutoff
and the the upper hybrid resonance before reaching the O-mode cutoff.
At higher frequency, above the maximum plasma frequency $\omega_{pe0}$ 
(calculated using $n_{e0}$), no O-mode cutoff is present in the plasma, no 
O-X conversion occurs, and the technique is not applicable. 

Note that the O-mode is only sensitive to the $\omega = \omega_{pe}$ cutoff, 
and the X-mode is only sensitive to the left and right cutoff and 
to the upper hybrid resonance. The technique relies on the 
oblique injection of an O-mode, and its partial conversion in a slow X-mode. 
The mode conversion occurs at the evanescent layer between the 
$\omega = \omega_{pe}$ and the $\omega = \omega_{L}$ locations. 
In practice, though, the injected polarization is rarely a pure O-mode, 
implying that some amount of fast X-mode is also injected. 
As a consequence, 
all cutoffs and resonances need to be taken into account in interpreting 
the complex wave-fields that will be discussed in the next Section.

\section{Simulation results}

\begin{figure}
  \begin{minipage}[b]{0.49\textwidth}\small $a)$\hfill$|E|$\hfill$f=10\,GHz$\hfill$\theta=30^o$\hfill$\,$\\ \includegraphics[width=6cm]{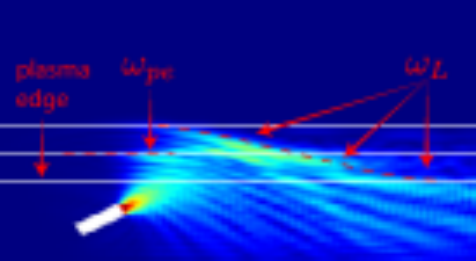} \end{minipage}
      \hfill
  \begin{minipage}[b]{0.49\textwidth}\small $b)$\hfill$|E|$\hfill$f=20\,GHz$\hfill$\theta=50^o$\hfill$\,$\\ \includegraphics[width=6cm]{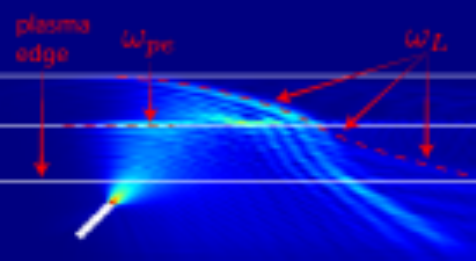} \end{minipage} \caption{
  Magnitude of the total electric field for (a) $f=10\,GHz$ and
  $\theta=30^o$ and (b) $f=20\,GHz$ and $\theta=50^o$. 
  The three horizontal white lines
  represent the plasma edge, the O-mode cutoff ($\omega=\omega_{pe}$)
  and the location where $\omega=\omega_{L}|_{\theta=90}$. The simulation domain
  extends $1\,m$ and $0.4\,m$ in the $\hat z$ and
  $\hat x$ directions, respectively.}  \label{fig:fwrf}
\end{figure}

In our study we investigated the wave launch, propagation and
reflection at the cutoff layers for different values of the injection
angle $\theta$ and of the wave frequency $f=\omega/2 \pi$. Because of
limitations in the present computational resources, in this initial 
paper we studied frequencies, 10-20 GHz, 
reflected at the very edge of the plasma. 
Extension to the 30-75 GHz range, corresponding to cutoff densities  
$n_e$=1-7$\times 10^{19}$m$^{-3}$, and thus more relevant to a tokamak pedestal, 
is left as a future work. 

Fig. \ref{fig:fwrf} shows contours of the magnitude $|\bf{E}|$ of the wave 
electric field for $f=10\,GHz$ and $\theta=30^o$ and
$f=20\,GHz$ and $\theta=50^o$. In our model the incident wave power is
distributed uniformly at the antenna mouth, with a wave field
polarization as from Eq. \ref{eq:pol}. After propagating in vacuum, the
waves reach the plasma edge ($x=0$, $n_e=0$). From there, the O-mode and
X-mode (if any) propagate until they reach, respectively, 
the $\omega=\omega_{pe}$ and the outermost between 
the $\omega=\omega_{R}$ and $\omega=\omega_{L}$ layer. 
As discussed before, at the low frequencies examined here,  
only the $\omega=\omega_{L}$ layer exists in the plasma 
(Figs.\ref{fig:res_cut},\ref{fig:fwrf}), but note that the actual 
turning point can lie slightly past it \cite{Volpe2003}.  
In the region between the 
O-mode and X-mode cutoff, only the X-mode wave can propagate. In these
figures, ripples in the wave fields magnitude indicate wave
interference which is symptomatic of wave reflection at the
aforementioned layers.

\begin{figure}
  \begin{minipage}[b]{\linewidth}\small $a)$\hfill$|E(x=0)|$\hfill$f=10\,GHz$\hfill$\,$\\\includegraphics[width=7cm]{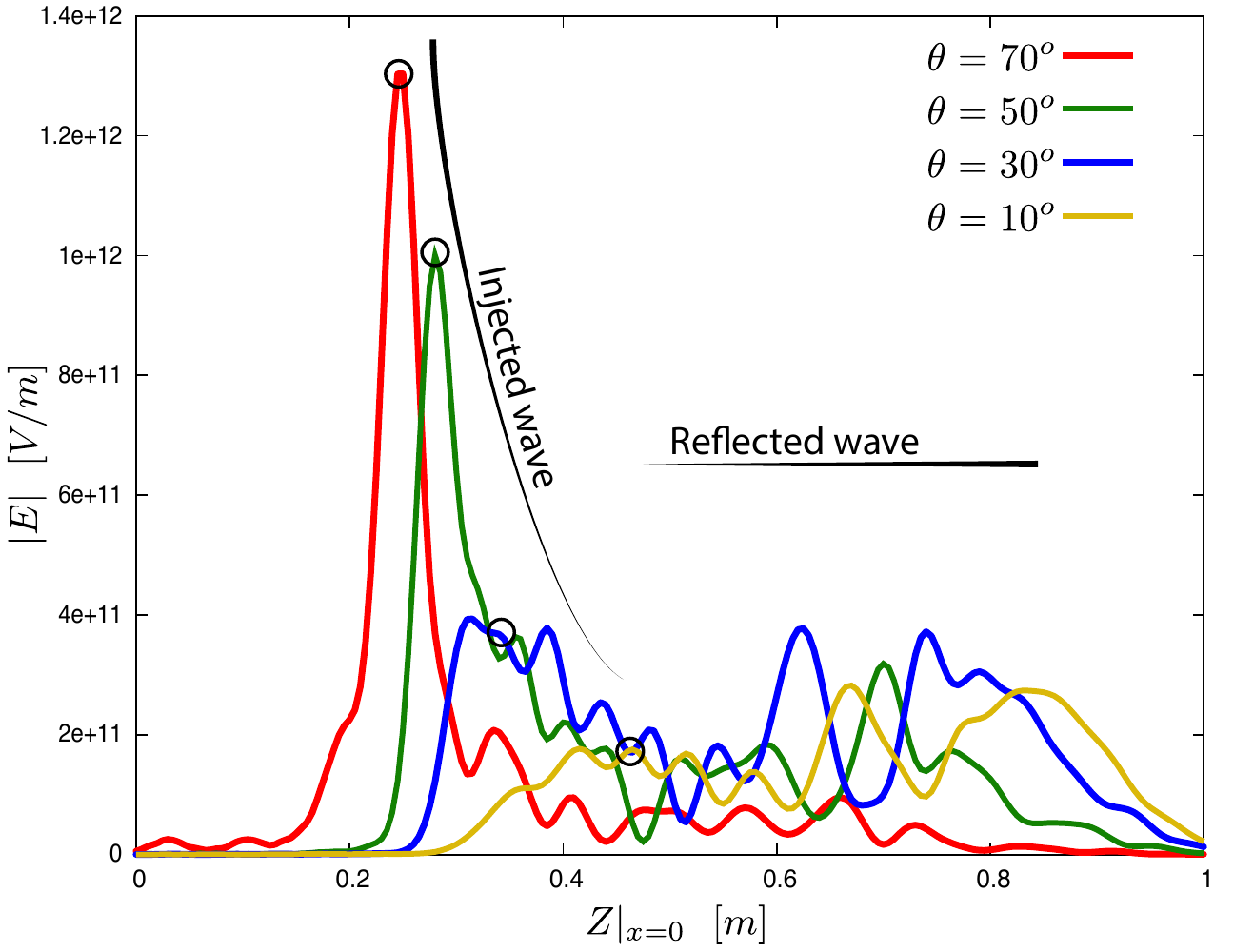} \end{minipage}\\
\begin{minipage}[b]{\linewidth}\small $b)$\hfill$|E(x=0)|$\hfill$f=20\,GHz$\hfill$\,$\\ \includegraphics[width=7cm]{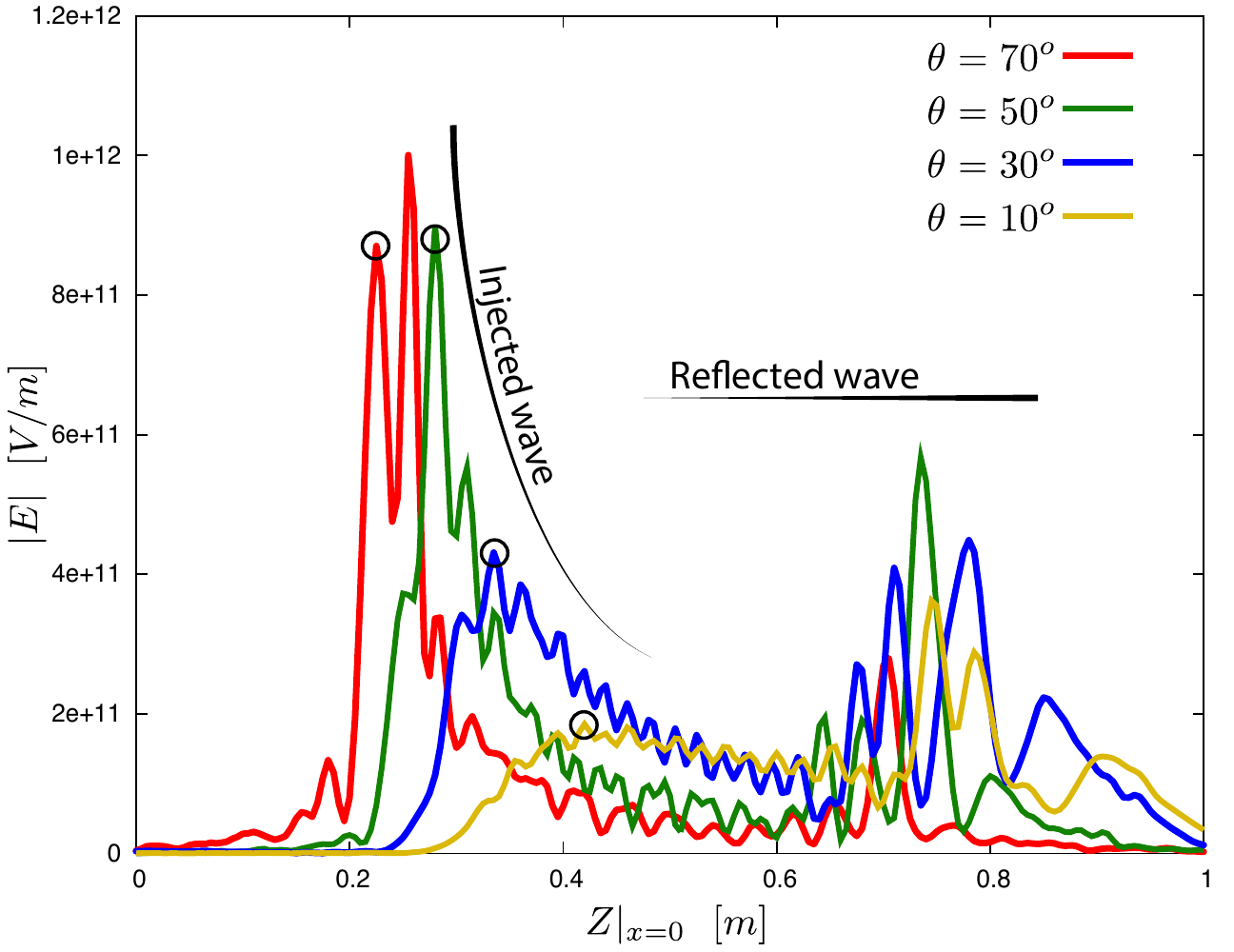} \end{minipage} \caption{Magnitude
  of the RF wave electric field computed along the plasma edge ($x=0$)
  for $\theta=70^o,50^o,30^o,10^o$ at $f= 10\,GHz$ and $20\,GHz$. The
  electric field for $x<0.4\,m$ is mostly representative of the waves
  launched by the antenna, while the values for $x>0.4\,m$ are mostly
  affected by the waves which are reflected at both
  $\omega=\omega_{pe}$ and $\omega=\omega_{L}$
  layers.}  \label{fig:cut}
\end{figure}

Figure \ref{fig:cut} shows the magnitude of the wave electric
field, as computed along the plasma edge ($x=0$) for various 
$\theta$ and $f$. This simulation is representative of what an array of 
receivers could measure, as a function of $z$. 
A peak is detected at $z<$0.4 m  
(that is, close to the launcher), especially for large values of $\theta$,  
when launch is nearly perpendicular to the plasma. 
This peak is partly due to waves that, from the launcher, directly 
reach the receivers without any interaction (reflection, mode conversion) 
with the O-mode cutoff. To avoid this, 
in the actual diagnostic the receivers will be placed in a retracted position, 
at larger minor radii compared with the launching antenna. 

The wave fields shown in the ranges of $0.4<z<1.0\,m$ are
mostly representative of the reflected waves at both
$\omega=\omega_{pe}$ and $\omega=\omega_{L}$ layers. As mentioned above, 
there is no right-hand cutoff layer in the plasma for 
$f=10\,GHz$ and $20\,GHz$, and reflection of the X-mode is due to the L cutoff. 
%
%

\begin{figure}
  \begin{minipage}[b]{\linewidth}\small $a) ~ ~
  E_x/1.00\times10^{6}\,[V/m]$\hfill$\,$\\ \includegraphics[width=7cm,clip=true,trim=0 0 0 20]{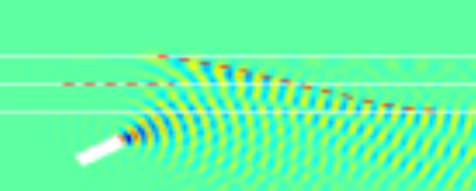} \end{minipage}
  \begin{minipage}[b]{\linewidth}\small $b) ~ ~
  E_y/0.77\times10^{6}\,[V/m]$\hfill$\,$\\ \includegraphics[width=7cm,clip=true,trim=0 0 0 20]{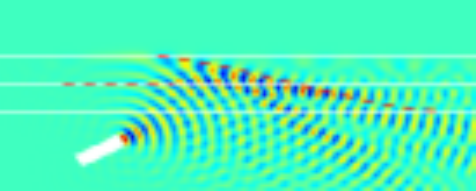} \end{minipage}
  \begin{minipage}[b]{\linewidth}\small $c) ~ ~
  E_z/0.81\times10^{6}\,[V/m]$\hfill$\,$\\ \includegraphics[width=7cm,clip=true,trim=0 0 0 20]{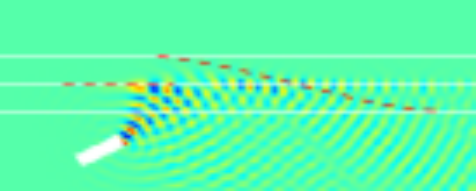} \end{minipage} \caption{Individual
  components of the waves total electric field $a)$ $E_x$ ; $b)$ $E_y$
  ; $c)$ $E_z$ for $f =10\,GHz$ at $\theta=30^o$. Absence of the $E_z$
  component indicates X-mode propagation. From this figure it is clear
  that the wave fields at the plasma edge ($x=0$) are a superposition
  of the reflections occurring at the O-mode and L-mode cutoff
  layers} \label{fig:Exyz}
\end{figure}

Figure \ref{fig:Exyz} illustrates more details for the propagation and
reflection of each component of the incident RF wave electric
field for the case of $f =10\, GHz$ at $\theta=30^o$. The O-mode
($E_z$) propagates in the plasma, and then is cutoff at
$\omega=\omega_{pe}$ and reflected back to the
plasma boundary. The X-mode ($E_y$) propagates up to the left cutoff
layer and is then reflected towards the plasma boundary. Similar
results were obtained for $f=20\, GHz$ at $\theta=50^o$. 

These COMSOL calculations in simplified
slab plasma geometry take only few minutes on a desktop computer.
For higher frequencies (thus, shorter wavelengths), the mesh needs to be 
refined, resulting in longer cpu times. 
The main limitation, however, is posed by the large memory requirements.

\section{Discussion, Conclusions and Future Work}
This study offers a preliminary outlook towards the possibility of 
reflectometrically measuring the magnetic field $\bf B$ 
at the O-mode cutoff layer. 
As discussed in connection with Eq.\ref{eq:refl} and 
Fig.\ref{fig:cont}, the information on the magnetic pitch angle is contained 
in the inclination of the angular contours of reflectivity, and the 
information on $|{\bf B}|$ is contained in the special direction yielding 
maximum OX conversion efficiency, hence minimum reflectivity of the O-mode. 
Full-wave simulations exhibited the expected ``hole'' in 
reflectivity for that special direction of incidence. 
At the frequencies analyzed in our study (10-20 GHz), simulations show 
also that this reflectivity hole is 
complicated by reflections at the X-mode cutoff
layer, which is itself a function of the the angle of incidence.

For frequencies higher than the electron cyclotron frequency as evaluated at 
the outer boundary, 
$\omega > \Omega_e (x=0)$, the right-hand cutoff and upper hybrid (UH) 
resonance will be present 
in the plasma. The cutoff will reflect power accidentally injected in the 
X-mode. Furthermore, the UH resonance and the evanescent layer lying between 
it and the R cutoff will nearly entirely prevent the X-mode generated by the 
OX conversion from reaching the receivers outside the plasma. 
In other words, the R cutoff and UH resonance 
would ``sanitize'' the diagnostic scheme from fast X and slow X-mode 
waves, respectively, and the receivers would only (or mostly) detect  
the reflected O-mode of interest.

In that case, wave propagation shares 
similarities with EBW emission diagnostics based on the BXO mode conversion. 
Yet, the anti-radar diagnostic will offer 
the advantages of being an active diagnostic and not relying on
the plasma to be over-dense and to be an intense EBW emitter with good 
B-X conversion efficiency. 
Future simulations at high frequency will 
be necessary to confirm that the X-mode wave does not affect the
signal at the receivers and will further strengthen confidence in the
feasibility of the diagnostic.

Further improvements could include extension 
to 3D, inclusion of the full details of the tokamak
magnetic field topology and a more realistic antenna
geometry. It is also planned to investigate the effects of toroidal ripples, 
density and magnetic fluctuations on reflectivity. Such
realistic models will provide a basis for the interpretation of the
experimental measurements.

\section*{Acknowledgements}
 This work was supported in part by the U.S. Department of Energy under DE-FG02-95ER54309.

\section*{Bibliography}

%

\end{document}